\documentclass[final,3p,times,twocolumn]{elsarticle}
\usepackage{amssymb}
\usepackage{amsmath}
\usepackage{lineno}
\usepackage{graphicx}

\journal{Physica C}

\begin{document}

\begin{frontmatter}

\title{Robust $s_{\pm}$-wave superconductivity against multi-impurity in iron-based superconductors }

\author[huang,Ren1,Ren2]{H. X. Huang\corref{cor1}}
%\ead{hxhuang@t.shu.edu.cn}
\author[huang]{S. C. Zhang}
\author[Gao]{Y. Gao}
\author[Ren1,huang,Ren2]{W. Ren}

\cortext[cor1]{Corresponding author. hxhuang@t.shu.edu.cn}

\address[huang]{Department of Physics, Shanghai University, Shanghai, 200444, China}

\address[Ren1]{Shanghai Key Laboratory of High Temperature Superconductors, Shanghai University, Shanghai, 200444, China}

\address[Ren2]{International Center of Quantum and Molecular Structures, Shanghai University, Shanghai, 200444, China}

\address[Gao]{Department of Physics and Institute of Theoretical Physics,
Nanjing Normal University, Nanjing, Jiangsu, 210023, China}

\begin{abstract}
Effects of disorder on electron-doped iron pnictides are investigated systematically based on self-consistent Bogoliubov-de Gennes equations. Multiply impurities with same scattering potential (SP) are randomly distributed in a square lattice. Probability distribution functions of normalized order parameters for different impurity concentrations $\delta_{imp}$, different electron doping concentrations $\delta$ are investigated for given SPs. Samples are found to be very robust against weak SP, in which order parameters do not have qualitative change even at very large $\delta_{imp}$. While strong SP is able to easily break down the order parameters. For moderate SP, variations of order parameters on and around impurities strongly depend on $\delta$, however the distribution functions of normalized order parameters have similar behavior as $\delta_{imp}$ increases. Compared with superconducting (SC) order, the magnetic order is more sensitive to multi-impurity effect. The spatial spin density wave pattern has already been destroyed before the system loses its superconductivity. Dependence of SC order on temperature is similar to that of impurity-free case, with the critical temperature being remarkably suppressed for high $\delta_{imp}$.
\end{abstract}
%\pacs{74.70.Xa, 74.62.En}

\end{frontmatter}

\section{introduction}
The discovery of the new family of layered iron-based SC materials has
attracted much attention since their discovery.~\cite{kam} The
parent compounds of iron pnictides are bad metal at low
temperatures and exhibit a spin-density-wave (SDW) order. Upon doping either electrons or holes into the system,
SDW will be suppressed and superconductivity emerges. Disorder thus cannot be avoided in the crystal growth procedure with intentional doping.  By substituting the iron ions with transition-metal (TM) elements Co or Ni,~\cite{jiun,hw,kn,zhou2} superconductivity can be realized and the dopant ions act as impurities at the same time.

Anderson's theorem tells us that the $s$-wave superconductivity is insensitive to small SP impurity. While in cuprates, impurity effect on the local density of states has been used as a prominent feature of $d$-wave  superconductor.~\cite{txiang,akem,pan} For iron-based superconductors,
impurity scattering effect has  been theoretically studied intensively~\cite{chub,bang,park,seng,voro,tsai,zhou1}, but mainly by considering a single impurity.

In TM doped cases, doping density is expected to be equal to impurity concentration.
X-ray absorption experiment~\cite{emb} and angle-resolved photoemission spectroscopy experiment~\cite{chang} show that TM substitution effect as well as  multi-impurity effect are rich in physics.
Experiments have measured the variation of superfluid density as function of temperature in crystal of $\mathrm{Ba(Fe_{1-x}Co_x)_2As_2}$ and in film $\mathrm{Ba(Fe_{1-x}Co_x)_2As_2}$.
The single crystal ~\cite{lan} shows a BCS-like behavior, while the epitaxial film ~\cite{jiey} has a wide range of linear dependance over temperature.
The difference between the crystal and film samples is usually considered that films are dirtier with large disorder.

Impurity effects are sensitive to SC order symmetry. One commonly believed idea is that the SC order is formed via spin fluctuation, leading to $s_{\pm}$-wave~\cite{kku,avc,sly,ygepl} pairing symmetry in iron pnictide. In this paper we address the robustness of $s_{\pm}$-wave superconductors against multi-impurity and only concern electron doped cases.

By employing a phenomenological model based on Bogoliubov-de Gennes (BdG) equations, we study systematically the multi-impurity effect in iron pnictides. The impurity Hamiltonian is described by an on-site point-like $\delta$-function potential, with SP strength $U_{imp}$ ranging from weak to strong. Multiple impurities have interference effect that can smear out the effect of single impurity. Although order parameters may be suppressed or enhanced on and around impurities, average values of order parameters decrease as $\delta_{imp}$ increases. And as a whole lattice they obey a statistical distribution function. We will study the probability distribution functions of normalized order parameters for different $\delta_{imp}$, different $\delta$ and different $U_{imp}$. With increasing $\delta_{imp}$, the height of the peak of distribution functions falls down at the beginning and then goes up for more larger $\delta_{imp}$ corresponding to the vanished order.

For weak SP, average values of the order parameters do not have qualitative change from that of the impurity-free case even when $\delta_{imp}$ is very large, suggesting that the system is very robust against weak multi-impurity. For moderate SP, SC order has finite values at $\delta_{imp}=0.2$, however, the stripe-like magnetic structure is destroyed. Strong SP breaks down the order parameters, for $U_{imp}=5.0$, the SDW pattern immediately disappear at $\delta_{imp}=0.0125$.

The rest of the paper is organized as follows. In Sec.~\ref{SEC:model}, we introduce the model. In Sec.~\ref{SEC:1}, we investigate the small SP cases, and in Sec.~\ref{SEC:2} and Sec.~\ref{SEC:3}
we discuss the moderate and strong SP cases respectively. Finally, we provide a summary and conclusion.

\section{model and formalism}\label{SEC:model}
We use a two-orbital four-band
phenomenological model~\cite{zhou1,zhang,zho,gao1,gao2,huang,huang1,huang2} to describe the Hamiltonian of iron pnictide. This model takes the $d_{xz}$ and $d_{yz}$ orbitals of the Fe ions into account and each unit cell accommodates two inequivalent Fe ions. It reproduces qualitatively the energy band structure observed by experiments~\cite{14,15,16,17,18,19}. The obtained results are consistent with the ARPES~\cite{arpes}, neutron scattering~\cite{ns} experiments.

The total Hamiltonian is defined as $H=H_{BCS}+H_{int}+H_{imp}$. Here $H_{BCS}$ is the BCS-like Hamiltonian, which includes the hopping term and the pairing term. The interaction Hamiltonian $H_{int}$ includes the on-site Coulomb interaction $U$ and Hund's coupling $J_{H}$ at the mean-field level. They are described by ~\cite{zhang,zho,gao1,gao2,huang,huang1,huang2}

\begin{subequations}
\begin{align}
H_{BCS}&=-\sum_{{\bf i}\nu^{\prime} {\bf j}\nu\sigma}(t_{{\bf i}\nu^{\prime}
            {\bf j}\nu}c^\dagger_{{\bf i}\nu^{\prime}\sigma}c_{{\bf j}\nu\sigma}+h.c.)-\mu\sum_{{\bf i}\nu^{\prime}\sigma}c^{\dagger}_{{\bf i}\nu^{\prime}\sigma}c_{{\bf i}\mu\sigma}
        \nonumber\\&\qquad+\sum_{{\bf i}\nu^{\prime} {\bf j}\nu\sigma}(\Delta_{{\bf i}\nu^{\prime}
        {\bf j}\nu}c^\dagger_{{\bf i}\nu^{\prime}\sigma}c^{\dagger}_{{\bf j}\nu\bar{\sigma}}+h.c.)\\
H_{int}&=U\sum_{{\bf i},\nu,\sigma\neq\bar{\sigma}}\langle
         n_{{\bf i}\nu\bar{\sigma}}\rangle n_{{\bf
         i}\nu\sigma}+(U-3J_H)\sum_{{\bf i},\nu^{\prime}\neq\nu,\sigma} \langle
          n_{{\bf i}\nu^{\prime}\sigma}\rangle n_{{\bf i}\nu\sigma}
          \nonumber\\
        &\qquad+(U-2J_H)\sum_{{\bf
i},\nu^{\prime}\neq\nu,\sigma\neq\bar{\sigma}} \langle n_{{\bf
i}\nu^{\prime}\bar{\sigma}}\rangle n_{{\bf i}\nu{\sigma}}.
\end{align}
\end{subequations}
In the above equations $n_{{\bf i}\nu\sigma}$ is the electron density operator at site ${\bf
i}$, orbital $\nu$, and spin $\sigma$, and $\mu$ is the chemical potential. $t_1$
represents the nearest-neighbor (nn) hopping between the same orbitals on Fe ions, $t_{2}$($t_{3}$) denotes the next-nearest-neighbor
(nnn) hopping between the same orbitals mediated by the up (down) As ions, $t_4$ is the nnn hopping between different
orbitals. See a schematic illustration in~\cite{huang}. We adopted the hopping parameters as $t_{1,2,3,4}=1,0.4,-2.0,0.04$, respectively. The energy and
SP are measured in unit of $t_1$.
The pairing order $\Delta_{{\bf i} {\bf j} }=\Delta_{{\bf i}\nu^{\prime} {\bf j}\nu}= \frac{V_{sc}}{2}\langle
c_{\bf{i}\nu^{\prime}\uparrow}c_{\bf{j}\nu\downarrow}-c_{\bf{i}\nu^{\prime}\downarrow}c_{\bf{j}\nu\uparrow}\rangle$ is chosen as the nnn
intra-orbital pairing with $\nu^{\prime}=\nu$ and $V_{sc}=1.5$. This kind
of pairing is consistent with the
$s_{\pm}$-wave superconductivity ~\cite{kku,14,mazin,yao,fawang} and has been widely used in
previous studies~\cite{zho,gao1,gao2,huang,huang1,huang2}.

Impurity Hamiltonian $H_{imp}$ is given by
\begin{equation}
H_{imp}=\sum_{{\bf m}\nu\sigma}U_{imp} c^{\dagger}_{{\bf
m}\nu\sigma}c_{{\bf m}\nu\sigma},
\end{equation}
where $\bf m$ is the impurity site acting as a scattering center with $U_{imp}$ varying from weak to strong. Our numerical calculation is performed on a $40\times 40$ square lattice with the periodic boundary conditions. For fixed $\delta_{imp}$, $\delta$ and $U_{imp}$ we self-consistently solve the BdG equation\cite{zhou1,huang} to obtain SC order $\Delta_{\bf{i}}=\frac{1}{4}\sum_{\tau}\Delta_{\bf{i,i+\tau}}$ and magnetic order $m_{\bf i}= \frac{1}{2}\sum_{\nu}(n_{{\bf
i}\nu\uparrow}-n_{{\bf i}\nu\downarrow})$, where $\tau$ is the four nnn diagonal links from site $\bf{i}$. In the absence of impurities, $m_{\bf{i}}$ shows a well-known stripe-like antiferromagnetic structure with two nearest rows having opposite $m_{\bf{i}}$.

In the calculations, the positions of the multiple impurities are spatially random. In order to obtain reliable probability distribution functions, for a fixed concentration $\delta_{imp}$, we carry out several numerical calculations of different configurations. The probability distribution functions of SC order $\rho_{\Delta}$ and magnetic order $\rho_{|m|}$ are then statistically analyzed. $\rho_{\Delta}$($\rho_{|m|}$) is denoted as the probability of $\Delta_{\bf{i}}/\Delta_0$($|m_{\bf{i}}|/|m_0|$), and here we define $\Delta_0$ and $|m_0|$ as the order parameters in the corresponding impurity-free case.
Summation of $\rho_{\Delta}$ and $\rho_{|m|}$ are required to be unity for renormalization condition. In order to convergence of our calculations we have verified that further sampling almost does not change $\rho_{\Delta}$ and $\rho_{|m|}$.

\section{weak scattering potential cases}\label{SEC:1}
\begin{figure}
\centering
      \includegraphics[width=8cm]{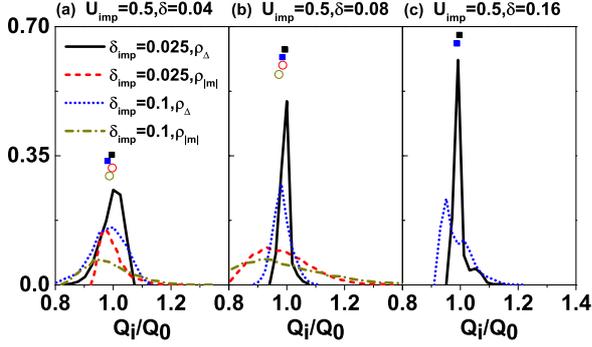}
\caption{(color online) Plots of $\rho_{\Delta}$ and $\rho_{|m|}$ for $U_{imp}=0.5$ and $\delta_{imp}=0.025, 0.1$ . Panels (a),(b),(c) correspond to $\delta=0.04,0.08,0.16$, respectively. The $x$ coordinate of the black(blue) solid square in each panel denotes the average value of $\bar{\Delta}_i$ for $\delta_{imp}=0.025$($0.1$) at the corresponding $\delta$. And that of the red(dark yellow) empty circle denotes the average value of $\bar{| m_i|}$ for $\delta_{imp}=0.025$($0.1$).  }\label{f1}
\end{figure}

For weak SP, we choose typical doping concentrations to investigate the distribution of the order parameters.
In the figures of distribution function, the horizontal axis denotes the value of the normalized order parameters $Q_{\bf{i}}/Q_0$, and the vertical axis denotes the corresponding probability $\rho_Q$, where $Q$ stands for $\Delta$ or $|m|$.

Fig.\ref{f1} is for  $U_{imp}=0.5$ with $\delta_{imp}$ chosen as $0.025$ and $0.1$. For $\delta_{imp}=0.025$, the peak of $\rho_{\Delta}$ is located at $\Delta_{\bf{i}}/\Delta_0=1.0$, despite the value
of $\delta$. It means that the most probable value of SC order is $\Delta_0$. Black solid line in Fig.\ref{f1}(a) shows that for underdoped case $\delta=0.04$, the value
of $\Delta_{\bf{i}}/\Delta_0$ is in the range $[0.85,1.07]$; for larger $\delta=0.08$, a narrower range can be seen in Fig.\ref{f1}(b). Panel(c) plot the overdoped cases $\delta=0.16$, in which the magnetic order vanishes and the peak of $\rho_{\Delta}$ is sharper. Increase $\delta_{imp}$ to $0.1$, the corresponding $\rho_{\Delta}$ becomes wider with lower peaks which are plotted by the blue dotted lines in Fig.\ref{f1}.

Magnetic order is more sensitive to multi-impurity, the peak of distribution function $\rho_{|m|}$ is lower than that of corresponding $\rho_{\Delta}$. As $\delta_{imp}=0.025$, for $\delta=0.04$ the range of $|m_{\bf{i}}|/|m_0|$ is $[0.95,1.15]$, for large doping level $\delta=0.08$ it is widely extended to $[0.8,1.2]$ as shown by the red dashed lines in Fig.\ref{f1}(a)(b). For $\delta_{imp}=0.1$, the corresponding $\rho_{|m|}$ is becomes wider than that of $\delta_{imp}=0.025$ which are shown by the dark yellow dash-dotted lines in Fig.\ref{f1}.

$x$ coordinate of the square(circle) dots in Fig.\ref{f1} is the average value of $\bar{\Delta}_i$($\bar{| m_i|}$) in the corresponding case. As can be seen $\bar{Q}_{\bf{i}}$ are close to $Q_0$ in all cases. Thus the system is very robust against weak SP regardless of $\delta_{imp}$. Even when $\delta_{imp}=0.2$ (not shown here), $\bar{Q}_{\bf{i}}$ does not have qualitative change with respect to $Q_0$. From Fig.\ref{f1}(c), we note that for large $\delta$, $\rho_{\Delta}$ has a small hump at the right side of $\Delta_{\bf{i}}/\Delta_0=1.0$. It is understandable since around the impurities $\Delta_{\bf{i}}$ are enhanced for high $\delta$.

\section{moderate scattering potential cases}\label{SEC:2}

\begin{figure}
\centering
      \includegraphics[width=8cm]{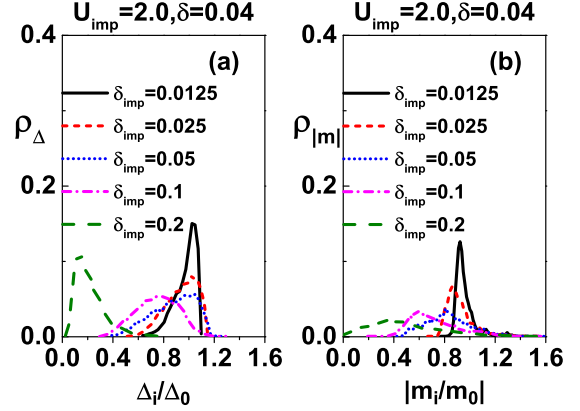}
\caption{(color online) (a)Plots of $\rho_{\Delta}$ as a function of $\Delta_{\bf{i}}/\Delta_0$ for different $\delta_{imp}$, at a fixed set of $\delta=0.04, U_{imp}=2.0$. (b) Similar to (a) but for $\rho_{|m|}$. From right to left the black solid, red short dashed, blue short dotted, pink dash-dotted, and green dashed lines represent $\delta_{imp}=0.0125,0.025,0.05,0.1,0.2$, respectively.  }\label{f3}
\end{figure}

\begin{figure}
\centering
      \includegraphics[width=8cm]{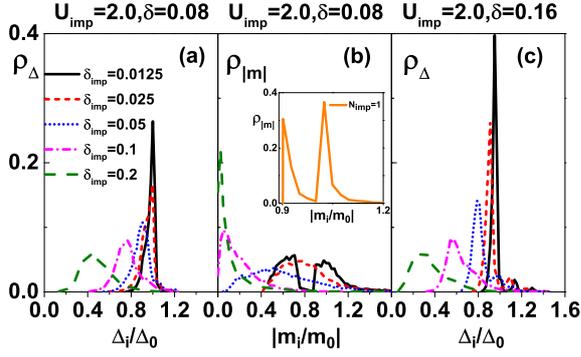}
\caption{(color online) (a)Plots of $\rho_{\Delta}$ as a function of $\Delta_{\bf{i}}/\Delta_0$ for different $\delta_{imp}$, at a fixed set of $\delta=0.08, U_{imp}=2.0$. (b) Similar to panel a but for $\rho_{|m|}$. Inset of (b) plot $\rho_{|m|}$ for $N_{imp}=1$. (c)Plots of $\rho_{\Delta}$ as a function of $\Delta_{\bf{i}}/\Delta_0$ for different $\delta_{imp}$, at a fixed set of $\delta=0.16, U_{imp}=2.0$. From right to left the black solid, red short-dashed, blue short-dotted, pink dash-dotted, and green dashed lines represent $\delta_{imp}=0.0125,0.025,0.05,0.1,0.2$, respectively.  }\label{f4}
\end{figure}

Then we investigate moderate SP cases, and Fig.\ref{f3} shows that for $U_{imp}=2.0$, the corresponding $\rho_{\Delta}$ and $\rho_{|m|}$ extend much wider than that of $U_{imp}=0.5$. As the impurity concentration $\delta_{imp}$ increases, the peak position moves to smaller value and the peak height falls down as seen in Fig.\ref{f3}(b) and all curves in Fig.\ref{f3}(a) except for the $\delta_{imp}=0.2$ case. For much big $\delta_{imp}$, $Q_{\bf{i}}$ on every site may be less than  $Q_0$, and the distribution will have a peak near zero with a considerable height. The green dashed curve in Fig.\ref{f3}(a) presents this regime with $\delta_{imp}=0.2$.

Distribution functions for $\delta=0.08$ have similar properties as that of $\delta=0.04$. Fig.\ref{f4}(a) shows that as $\delta_{imp}$ increases gradually from zero to $0.2$, the peak height of corresponding $\rho_{\Delta}$ decreases. Since no one site has $\Delta_i=0$, the system is in a stable SC phase for $\delta_{imp}=0.2, U_{imp}=2.0$. As for $\rho_{|m|}$, the height of the peak goes up when $\delta_{imp}\geq0.1$ which are shown in Fig.\ref{f4}(b). When $\rho_Q$ enters this regime, the corresponding order $Q$ tends to disappear. It is worth noting that different from $\delta=0.04$, see the black solid line in Fig.\ref{f4}(b), $\rho_{|m|}$ for $\delta=0.08$ has two humps instead of one sharp peak. The reason is that when $\delta$ near optimal doping, small amount of impurities can induce the lattice separate into two sublattice  since each unit cell contains two inequivalent Fe ions. This is consistent with previous study~\cite{zhou1}. For $\delta=0.08$, $\rho_{|m|}$ of single impurity is plotted in the inset of panel(b) with two pronounced peaks shown up. Multi-impurity will smear out the two humps. In order to see it clearly, we show the images of normalized magnetic order $|m_{\bf{i}}|/|m_0|$ for $\delta=0.04$ and $0.08$ in the following.

\begin{figure}
\centering
      \includegraphics[width=8cm]{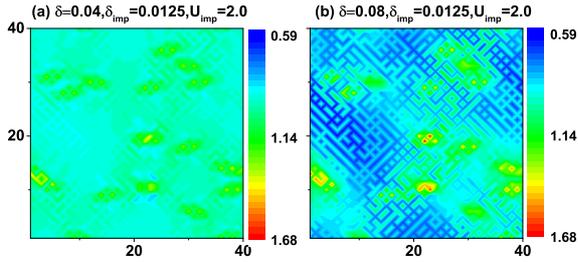}
\caption{(color online) Images of $|m_{\bf{i}}|/|m_0|$ for $\delta_{imp}=0.0125, U_{imp}=2.0$. Panel (a) is for $\delta=0.04$ and panel (b) is for $\delta=0.08$. }\label{f4.5}
\end{figure}

\begin{figure}
\centering
      \includegraphics[width=8cm]{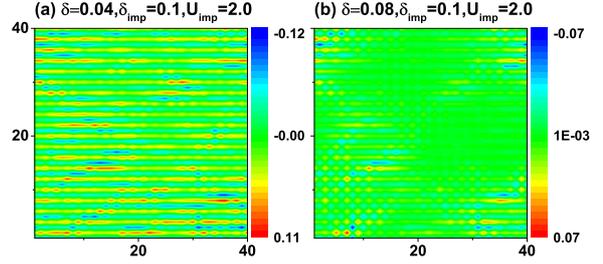}
\caption{(color online) Images of $m_{\bf{i}}$ for $\delta_{imp}=0.1, U_{imp}=2.0$. Panel (a) is for $\delta=0.04$ and panel (b) for $\delta=0.08$. }\label{f4.6}
\end{figure}

We can see form Fig.\ref{f4.5} that $|m_{\bf{i}}|$ at or around the impurities is enhanced for both doping concentrations. Panel (a) shows that for $\delta=0.04$, the main part of the lattice has $|m_{\bf{i}}|$ less but close to $|m_0|$ leading to the sharp peak in Fig.\ref{f3}(b). While Fig.\ref{f4.5}(b) shows that for $\delta=0.08$, except for the enhanced region of $|m_i|$, two sublattice crossed each other explicitly, which contribute to the two humps in Fig.\ref{f4}(b) for $\delta_{imp}=0.0125$.

Instead of $|m_{\bf{i}}|/|m_{\bf{0}}|$, Fig.\ref{f4.6} depicts $m_{\bf{i}}$ for $\delta_{imp}=0.1$, $U_{imp}=2.0$ cases. The stripe-like pattern is still stable for $\delta=0.04$ as shown in Fig.\ref{f4.6}(a). When the doping level increases to $\delta=0.08$, $m_{\bf{i}}=\pm0.04$ for the pure sample; however, the multi-impurity makes the stripe-like pattern almost disappear, since we have many sites $m_{\bf{i}}=0$ in Fig.\ref{f4.6}(b). Whereas magnetic order is more sensitive to multi-impurity in comparison to no one site has zero SC order.

\begin{figure}
\centering
     \includegraphics[width=8cm]{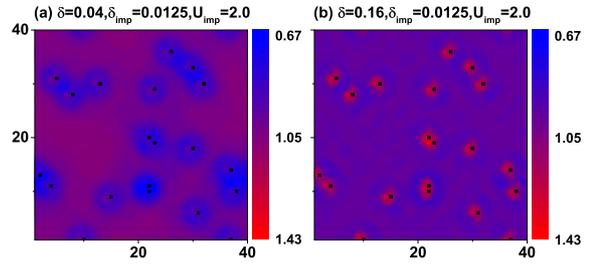}
\caption{(color online) Images of $\Delta_{\bf{i}}/\Delta_0$ for $\delta_{imp}=0.0125$, $U_{imp}=2.0$. Panel (a) is for $\delta=0.04$ and panel (b) for $\delta=0.16$. Where the black solid squares stand for the sites of the multiple impurities. }\label{f6}
\end{figure}

Results of overdoped $\delta=0.16$, moderate $U_{imp}=2.0$ are shown in Fig.\ref{f4}(c). The peak height of $\rho_{\Delta}$ is much higher than that of $\delta=0.08$ and $\delta=0.04$ especially for lower impurity concentration. The main feature is very similar to that of $\delta=0.08$ except that $\rho_{\Delta}$ can have satellite peaks located at $\frac{\Delta_{\bf{i}}}{\Delta_0}>1.0$ for $\delta_{imp}\leq0.05$. Thus in Fig.\ref{f6} we present the image of $\Delta_{\bf{i}}/\Delta_{\bf{0}}$ for $\delta=0.16$ and $\delta=0.04$ for comparison.

Fig.\ref{f6} shows that the multi-impurity effect on $\Delta_{\bf{i}}$ is significantly different for low doping level and high doping level. Panel (a) shows for underdoped $\delta=0.04$, around the impurities $\Delta_{\bf{i}}$ is suppressed. Away from the impurities $\Delta_i$ is close to $\Delta_0$ and corresponding to the sharp peak of Fig.\ref{f3}(a). On the contrary, for overdoping level $\delta=0.16$, on the impurity sites as well as on most of their nn sites $\Delta_{\bf{i}}$ are enhanced, which can be clearly seen in Fig.\ref{f6}(b). The sites of enhanced $\Delta_i$ lead to additional satellite peaks located at $\frac{\Delta_{\bf{i}}}{\Delta_0}>1.0$ in Fig.\ref{f4}(c). As $\delta_{imp}$ increases, interference between the multi-impurities will smear out these small peaks.

\begin{figure}
\centering
     \includegraphics[width=9cm]{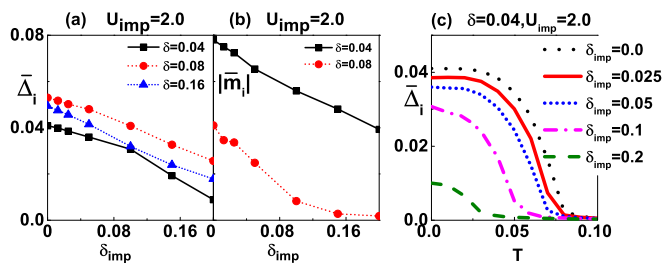}
\caption{(color online) (a)Schematic plot of $\bar{\Delta}_{\bf{i}}$ versus $\delta_{imp}$ for different $\delta$ at $U_{imp}=2.0$. (b) Schematic plot of $\bar{| m_i|}$ versus $\delta_{imp}$ for different $\delta$ at $U_{imp}=2.0$. (c) Plots of $\bar{\Delta}_{\bf{i}}$ as a function of $T$ for different $\delta_{imp}$ for fixed $\delta=0.04,U_{imp}=2.0$. }\label{f7}
\end{figure}

Although order parameters may be enhanced at some sites, the average value $\bar{Q}_i$ is less than the corresponding $Q_0$ for all cases. Fig.\ref{f7}(a) shows that $\bar{\Delta}_i$ decreases as $\delta_{imp}$ increases for all doping levels. The decreasing rate depends on $\delta$ as well as on $\delta_{imp}$. All $\bar{\Delta}_i$ have finite values at $\delta_{imp}=0.2$. Magnetic order $\bar{| m_i|}$ is also reduced as $\delta_{imp}$ increases as shown in Fig.\ref{f7}(b). For $\delta=0.08$, $\bar{| m_i|}$  has a small finite value at $\delta_{imp}=0.1$, while the SDW pattern is already largely broken for $\delta_{imp}=0.1$ as seen in Fig.\ref{f4.6}(b).

Fig.\ref{f7}(c) plots $\bar{\Delta}_{\bf{i}}$ as a function of $T$ for different $\delta_{imp}$ at fixed $U_{imp}=2.0$, and $\delta=0.04$. The black dotted line denotes the impurity-free case for comparison. From top to down the curves for increased $\delta_{imp}$ have similar shapes. At low $T$, the curves are flat, and the higher $\delta_{imp}$ the shorter the flat region. As $T$ increases, $\bar{\Delta}_i$ falls more quickly and vanishes at the critical temperature $T_c$.  At lower $\delta_{imp}$, $T_c$ is less deviated from the clean limit; while for higher $\delta_{imp}$, $T_c$ is remarkably suppressed.

\section{strong scattering potential cases}\label{SEC:3}

\begin{figure}
\centering
      \includegraphics[width=8cm]{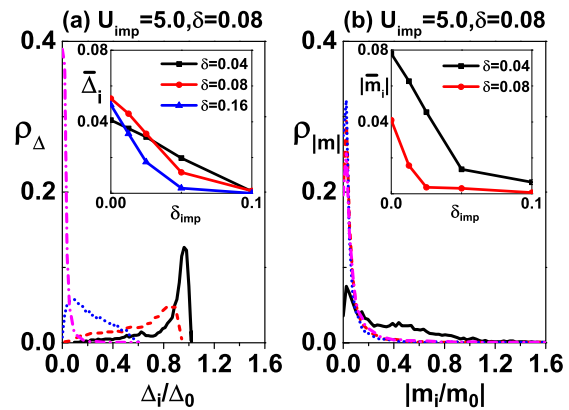}
\caption{(color online) (a)Plots of $\rho_{\Delta}$ as a function of $\Delta_{\bf{i}}/\Delta_0$ for different $\delta_{imp}$, at a fixed set of $\delta=0.08, U_{imp}=5.0$. (b) Similar to panel(a) but for $\rho_{|m|}$. From right to left the black solid, red short-dashed, blue short-dotted, pink dash-dotted represent $\delta_{imp}=0.0125,0.025,0.05,0.1$, respectively.  Inset of panel(a) plots $\bar{\Delta}_{\bf{i}}$ as function of $\delta_{imp}$ for different $\delta$ at $U_{imp}=5.0$. Inset of panel(b) plots $\bar{| m_i|}$  as function of $\delta_{imp}$ for different $\delta$ at $U_{imp}=5.0$. }\label{f8}
\end{figure}

Strong SP multi-impurity suppress the order parameters more easily, the relevant properties are depicted in Fig.\ref{f8} for large $U_{imp}=5.0$. We only present the results of $\delta=0.08$, since other doping cases have similar results. Panel(a) shows that when $\delta_{imp}\geq0.05$(blue short-dotted line), the order parameters are concentrated near zero with a very sharp peak for $\delta_{imp}=0.1$(pink dash-dotted line), and the system totally loses superconductivity. Magnetic order is even more sensitive to impurities, panel(b) shows that when $\delta_{imp}>0.0125$, sharp peaks of $\rho_{|m|}$ are all squeezed to zero with vanished magnetic order.

For all doping levels, average order parameters decrease more quickly as $\delta_{imp}$ increases which can be seen in the inset of Fig.\ref{f8}(a). It shows that at $\delta_{imp}=0.1$, we have $\bar{\Delta}_i=0.0$ for all cases and inset of Panel(b) shows that $\bar{|m_i|}$ decreases sharply as well. Although at $\delta=0.04$ and $\delta_{imp}=0.1$, $\bar{| m_i|}$ has a very small finite value, the SDW pattern is broken.

\section{summary}\label{SEC:summary}
Motivated by experiments of TM doped iron-based superconductors and the different superfluid densities between bulk and film samples, we investigate multi-impurity effect systematically based on a phenomenological model. $s_{\pm}$-wave superconductivity has its distinct properties when impurities are introduced. We use a number of configurations with random impurities to study the probability distributions of the order parameters.

For weak SP, averaged order parameters do not have qualitative change even at very high impurity concentration $\delta_{imp}$. They resemble the properties of the clean sample, just like by setting the Coulomb interaction $U$ slightly larger and the system is still in a stable SC phase.

For moderate SP, small $\delta_{imp}$ but larger $\delta$, the distribution functions of $\rho_{\Delta}$  have a sharp peak together with a small hump, since the vicinity of impurities consists a considerably main part with enhanced $\Delta_i$. As $\delta_{imp}$ increases, the hump disappears due to the interference between many impurities.

The multi-impurity effect on order parameters is remarkably different for low and high doping levels; however, the averaged order parameters are suppressed with the increasing $\delta_{imp}$ in all cases. The probability distribution functions have similar behavior as $\delta_{imp}$ increases for moderate and strong SPs. The distribution peaks move towards zero for the larger $\delta_{imp}$, meanwhile the peak near unity is suppressed and then replaced by the increased peak corresponding to a vanished order.

$s_{\pm}$-wave superconductor is robust again moderate SP multi-impurity effect, since $\bar{\Delta}_i$ is finite for all doping levels at large $\delta_{imp}=0.2$ and no site has $\Delta_i=0$. While magnetic order is more sensitive to multi-impurities. At $\delta=0.1$, although average $\bar{|m_i|}$ is finite, the stripe-like SDW pattern has already been destroyed. The $T_c$ values of the samples are robust against small $\delta_{imp}$, and are increasingly  suppressed  for the higher impurity concentration.

Finally, we find that the strong SP is able to easily break down the order parameters. All the above features from our calculations provide some deep understanding of the difference between clean and dirty samples of iron-based superconductors.

\section{acknowledgements}
This work was supported by the National Key Basic Research Program of China (Grant No. 2015CB921600), NSF of Shanghai (Grant No. 13ZR1415400), Shanghai Key Lab for Astrophysics(Grant No. SKLA1303), NSFC (Grants No. 11204138 and No. 11274222), NSF of Jiangsu Province of China (Grant No. BK2012450), the QiMingXing Project (Project No. 14QA1402000) of the Shanghai Municipal Science and Technology Commission, the Eastern Scholar Program, and the Shuguang Program (Grant No. 12SG34) from the Shanghai Municipal Education Commission.

%\bibliographystyle{elsarticle-num}

\end{document}